\documentclass[12pt]{article}
\usepackage{graphicx}
\begin{document}
\begin{center}
{\bf The heat engine of magnetic black holes in AdS space with rational nonlinear electrodynamics} \\
\vspace{5mm} S. I. Kruglov
\footnote{E-mail: kruglov@rogers.com}
\underline{}
\vspace{3mm}

\textit{Department of Physics, University of Toronto, \\60 St. Georges St.,
Toronto, ON M5S 1A7, Canada\\
Canadian Quantum Research Center, \\
204-3002 32 Ave., Vernon, BC V1T 2L7, Canada} \\
\vspace{5mm}
\end{center}
\begin{abstract}
The heat engine of magnetic black holes in Einstein-AdS gravity coupled to rational nonlinear electrodynamics, as the working substance, is studied. The dynamical negative  cosmological constant is considered as a thermodynamic pressure. We investigate the efficiency of black hole heat engines in extended space thermodynamics for rectangle closed path in the $P - V$ plane and the maximally efficient Carnot cycles. The exact efficiency formula which is written in terms of the mass of the black hole is obtained. It was demonstrated that the black hole efficiency decreases when the nonlinear electrodynamics coupling increases and the black hole efficiency increases if the magnetic charge increases. The relation between the efficiency, event horizon radiuses (entropy) and pressure is obtained. We study an efficiency of the holographic heat engine of a cycle in the vicinity of a critical point. Thus, the heat engine of our model can produce work.
\end{abstract}
\vspace{5mm}

\section{Introduction}

In extended phase space black hole thermodynamics the negative cosmological constant, which is the dynamical variable, is connected with the pressure, $P=-\Lambda/(8\pi)$ \cite{Kastor,Dolan1,Dolan2,Dolan3,Cvetic2}. Then the first law of black hole thermodynamics is formulated as
\begin{equation}
d M = TdS + VdP + \Phi dq,
\label{1}
\end{equation}
where the black hole mass $M$ is treated as an enthalpy, $V$ is the geometric volume $4\pi r^3_+/3$  ($r_+$ is the event horizon radius), $q$ is a charge, and $\Phi$ is a potential. When Einstein's gravity in Anti de Sitter (AdS) space couples to nonlinear electrodynamics (NED), the coupling $\beta$ is the thermodynamic variable conjugated to so called vacuum polarization \cite{Mann}. In this case the first law of black hole thermodynamics (1) is modified by adding the term ${\cal B}d\beta$ with ${\cal B}$ being the conjugate to coupling $\beta$.
The Einstein's gravity with a negative cosmological constant allows us to consider a holographic picture where a black hole is a system that is dual to the conformal field theory (CFT) \cite{Maldacena,Witten,Witten1}. The holography (AdS/CFT correspondence) can help us to solve nontrivial problems in condensed matter physics \cite{Kovtun,Kovtun1,Hartnoll}. The heat engine of a black hole as the working substance was studied firstly by Johnson \cite{Johnson}. It was shown that considering a closed path in the $P-V$ plane the useful mechanical work of static AdS black holes can be obtained from heat energy. For Born--Infeld electrodynamics in AdS space a black hole as a heat engine was studied in \cite{Johnson3}. The authors of Ref. \cite{Mann1} studied the effects of spacetime dimension, rotation, and higher curvature corrections on the efficiency of the holographic heat engine cycle. It is worth noting that the efficiency depends on the black hole (working substance) equation of state and the choice of the cycle. In \cite{Johnson4,Johnson7} a benchmarking scheme was described that allows to compare the efficiencies of black holes (as working substances) in heat engines for a circular cycle in the $P-V$ plane. Studies of holographic heat engines employing different black hole heat engines were considered in Refs. \cite{Johnson5,Pedraza,Johnson6,Zhang,Rosso,Yerra,Bamba,Johnson8,Ahmed,Yerra1,Moumni,Jafarzade,Ujjal} and in other papers.

This paper studies the heat engine of a magnetic black hole in Einstein-AdS gravity coupled to rational nonlinear electrodynamics (RNED) which was proposed in \cite{Kruglov} (see also \cite{Kruglov1}). It was proven by Bronnikov \cite{Bronnikov} that electrically charged black holes with the Lagrangian function ${\cal L}({\cal F})$, possessing the Maxwell limit at weak fields, do not have a
regular center for a static, spherically symmetric solution of general relativity. Therefore, we will study only magnetically charged black holes. In the case electrically charged black holes, thermodynamic behavior of black holes will be different because of equation of state. It is of interest also to study how quantum corrections can influence the thermodynamic properties and efficiency of the black hole heat engine. We will study this in further.
We consider also holographic heat engines of our model in the vicinity of a critical point. The black hole thermodynamics in extended phase space in the framework of Einstein-AdS-RNED was studied in \cite{Kruglov2}.
The Einstein–-Born–-Infeld solution, critical points and stability were investigated in \cite{Behnam}. In \cite{Behnam1} the effect of the non-perturbative correction on the thermodynamics quantities was studied.
The RNED is similar to Born--Infeld electrodynamics which is without singularities at the center and at weak-field limit it is converted into the Maxwell electrodynamics. But there are differences: the Born--Infeld Lagrangian at high field limit becomes infinite but for RNED one has $\mbox{lim}_{{\cal F}\rightarrow\infty}{\cal L}({\cal F})=-1/(8\pi\beta)$ which is finite. It was stated in \cite{Bronnikov2} that NED models with finite limit at high fields are preferable. The comparison of RNED with other NED models was done in \cite{Bronnikov2}.
In addition, the Born--Infeld  theory does not admit the birefringence phenomena which is predicted by QED that is now on the experimental verification. At the same time the Generalized Born--Infeld theory \cite{Kruglov5} (see also \cite{Kruglov6}) as well as RNED (with the additional therm $a(\textbf{E}\cdot\textbf{B})^2$ in RNED Lagrangian \cite{Kruglov7}) describes the birefringence effect. The gravity coupled to RNED describes the universe inflation \cite{Kruglov3} and reproduces the shadow of $M87^*$ black hole \cite{Kruglov4}. Here we use Einstein–AdS–RNED black hole as the working substance. We will show that within our model the heat engine of black holes can produce work.

The structure of the paper is as follows. In Section 2 we review the model of Einstein-AdS gravity coupled to RNED. The thermodynamic variables such as the metric function, black hole mass, Hawking temperature and pressure are given.
The black hole heat efficiency is studied in Section 3. A rectangle closed path in the $P-V$ plane and Carnot's cycle are considered. We have obtained and compared efficiencies for these two cycles at some model parameters. The holographic heat engine cycle in the vicinity of
a critical point for our model is studied in Section 4. Section 5 is the summery of results obtained.

We will employ geometrical units with $G=c=\hbar=k_B=1$.

\section{Black hole solution}

Einstein-AdS gravity coupled to RNED is given by the action \cite{Kruglov2}
\begin{equation}
I=\int d^{4}x\sqrt{-g}\left(\frac{R-2\Lambda}{16\pi}+\mathcal{L}(\mathcal{F}) \right),
\label{2}
\end{equation}
where $\Lambda=-3/l^2$, $l$ is the AdS radius and the RNED Lagrangian is given by
\begin{equation}
{\cal L}=-\frac{{\cal F}}{4\pi(1+2\beta {\cal F})}.
\label{3}
\end{equation}
${\cal F}=F^{\mu\nu}F_{\mu\nu}/4=(B^2-E^2)/2$, and $E$ and $B$ are the electric and magnetic fields. We will study magnetic black holes and, therefore, $E=0$, $B=q/r^2$ with $q$ being a magnetic charge.
The spherical symmetry solution for the metric function is
\begin{equation}
f(r)=1-\frac{2m(r)}{r}-\frac{\Lambda r^2}{3},
\label{4}
\end{equation}
where the mass function reads \cite{Kruglov2}
\[
m(r)=m_0+\frac{q^{3/2}g(r)}{8\sqrt{2}\beta^{1/4}},~~~~
g(r)\equiv \ln\left(\frac{r^2-\sqrt{2q}\beta^{1/4}r+q\sqrt{\beta}}{r^2+\sqrt{2q}\beta^{1/4}r+q\sqrt{\beta}}\right)
\]
\begin{equation}
+2\arctan\left(\frac{\sqrt{q}\beta^{1/4}+\sqrt{2}r}{\sqrt{q}\beta^{1/4}}\right)
-2\arctan\left(\frac{\sqrt{q}\beta^{1/4}-\sqrt{2}r}{\sqrt{q}\beta^{1/4}}\right)
\label{5}
\end{equation}
and $m_0$ is the Schwarzschild mass (the integration constant). The BH magnetic mass is given by
 \begin{equation}
m_M=4\pi\int_0^\infty \rho_M r^2dr=\frac{\pi q^{3/2}}{4\sqrt{2}\beta^{1/4}},~~~~\rho_M =\frac{q^2}{8\pi (r^4+q^2\beta)},
\label{6}
\end{equation}
where $\rho_M$ is the density of the magnetic energy.
From equation $f(r_+)=0$, where $r_+$ is the event horizon radius, and Eqs. (4) and (5) one finds the black hole mass
\begin{equation}
M(r_+)=\frac{r_+}{2}+\frac{r_+^3}{2l^2}-\frac{q^{3/2}g(r_+)}{8\sqrt{2}\beta^{1/4}}+\frac{\pi q^{3/2}}{4\sqrt{2}\beta^{1/4}}.
\label{7}
\end{equation}
From the definition of the Hawking temperature $T=f'(r)|_{r=r_+}/(4\pi)$ and Eqs. (4), (5), we obtain the Hawking temperature \cite{Kruglov2}
\begin{equation}
T=\frac{1}{4\pi}\biggl(\frac{1}{r_+}+\frac{3r_+}{l^2}-\frac{q^2r_+}{r_+^4+\beta q^2}\biggr),
\label{8}
\end{equation}
where $P=3/(8\pi l^2)$. From Eq. (8) we find Equation of State (EoS)
\begin{equation}
P=\frac{T}{2r_+}-\frac{1}{8\pi r_+^2}+\frac{q^2}{8\pi (r_+^4+\beta q^2)}.
\label{9}
\end{equation}
One can verify from Eqs. 7 and 8 that $V=\partial M/\partial P=4\pi r_+^3/3$ which is the geometric/thermodynamic volume of a black hole conjugated to the pressure $P$ and the Hawking--Bekenstein entropy is $S=\int_0^{r_+}(1/T)dM=\pi r_+^2$. The event horizon radius $r_+$, which is the higher root of equation $f(r_+)=0$, can be replaced by $S$ or $V $ because they are not independent.
It is worth noting that at $\beta=0$ the Hawking temperature (8) and pressure (9) become $T$ and $P$ corresponding to  Reissner–-Nordstr\"{o}m-AdS space. It should be noted that at great charge $q$ and small $\beta$ the Hawking temperature (8) becomes negative, i.e. nonphysical.

\section{The black hole heat efficiency}

A heat engine is defined as a closed path in the $P-V$ plane with $P(V,T)$ being EoS. The
input amount of heat is $Q_H$ and the exhaust amount is $Q_C$ with the total mechanical work done $W=Q_H-Q_C$.
Then the efficiency of the heat engine is given by
\begin{equation}
\eta=\frac{W}{Q_H}=1-\frac{Q_C}{Q_H}
\label{10}
\end{equation}
and it depends upon a closed path and EoS.  The heats $Q_H$ and $Q_C$ may be treated as going from heat baths of radiation around
black holes \cite{Hawking}.
In the classic maximally efficient Carnot cycle two isotherms are connected by two adiabatic paths while in the classic Stirling cycle two isochoric paths connect two temperatures. It was shown in \cite{Johnson} that for static black holes, the Carnot engine is also a Stirling engine and they are reversible because the total entropy flow is zero. For the static black holes the thermodynamic volume and the entropy are not independent and, therefore, that adiabats and isochores are the same for Carnot and Stirling cycles.
The Carnot efficiency is given by the temperature difference, $\eta_C=W/Q_H=1-T_C/T_H$. Here, a pair of isotherms are at temperatures $T_H$ and $T_C$, where in isothermal expansion some heat is absorbed at temperatures $T_H$, and in isothermal compression some heat gets out at the temperature $T_C$. The heat flow is shown in Fig. 1. The upper and lower isobars give the net flow of heat $Q_H$ and $Q_C$, correspondingly
\begin{figure}[h]
\includegraphics [height=2.7in,width=5in] {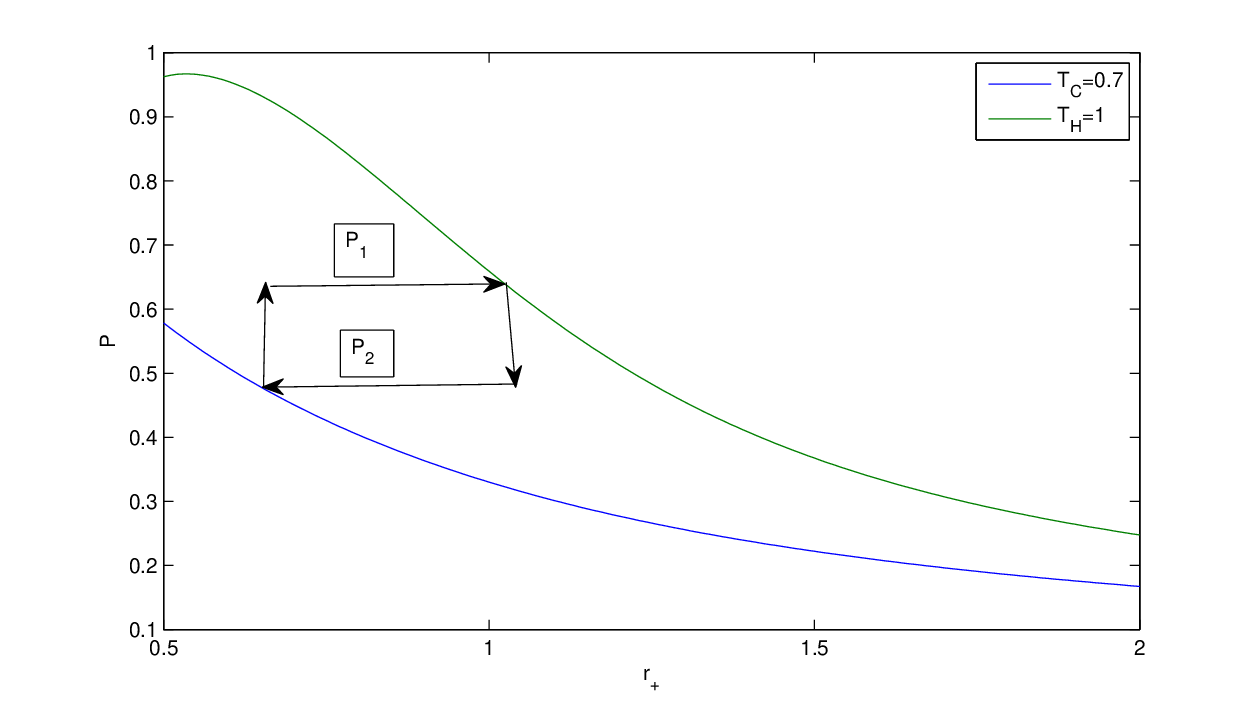}
\caption{\label{fig.1} A rectangle closed path in the $P - V$ plane and two isotherms with $T_C=0.7$, $T_H=1$, $\beta=1$, $q=1$, $r_+=\sqrt[3]{3V/(4\pi)}$.}
\end{figure}
\begin{equation}
Q_H=\int_{T_1}^{T_2} C_P(P_1,T)dT,~~~~Q_C=\int_{T_1}^{T_2} C_P(P_2,T)dT.
\label{11}
\end{equation}
It should be noted that there is not the heat exchange in the isochoric process because $C_V=0$ (processes corresponding to vertical arrows in Fig. 1). The heat capacity is given by
\begin{equation}
C_P=T\left(\frac{\partial S}{\partial T}\right)_{P,q}=\frac{T\partial S/\partial r_+}{\partial T/\partial r_+}=\frac{2\pi r_+ T}{\partial T/\partial r_+},
\label{12}
\end{equation}
where the black hole entropy $S=\pi r_+^2$. Making use of Eqs. (8), (11) and (12) we obtain
 \[
 Q_H=\frac{r_2-r_1}{2}+\frac{r_2^3-r_1^3}{2l_1^2}-\frac{q^{3/2}\left(g(r_2)-g(r_1)\right)}{8\sqrt{2}\beta^{1/4}},
 \]
\begin{equation}
Q_C=\frac{r_2-r_1}{2}+\frac{r_2^3-r_1^3}{2l_2^2}-\frac{q^{3/2}\left(g(r_2)-g(r_1)\right)}{8\sqrt{2}\beta^{1/4}},
\label{13}
\end{equation}
where $r_1=\sqrt[3]{3V_1/(4\pi)}$, $r_2=\sqrt[3]{3V_2/(4\pi)}$, $l_1^2=3/(8\pi P_1)$, $l_2^2=3/(8\pi P_2)$. It is worth mentioning that the efficiency of the heat engine (10) with $Q_H$ and $Q_C$ given in Eq. (13) can be written through the black hole mass (7) in accordance with the formula presented in \cite{Johnson1}. The useful work done in our holographic engine is $W=Q_H-Q_C=0.5(r^3_2-r^3_1)(1/l_1^2-1/l_2^2)$.
Thus to increase the engine useful work one can increase the difference $r_2-r_1$ or/and $\Lambda_2-\Lambda_1$ ($\Lambda=-3/l^2$) of cycles.
With the help of Eqs. (8), (10) and (13) we plotted in Fig. 2 the black hole heat efficiency $\eta$
vs. $\beta$ for a cycle with two isobars and the heat efficiency  $\eta_C$ for a cycle with two isotherms for particular parameters, as an example.
\begin{figure}[h]
\includegraphics [height=2.7in,width=5in] {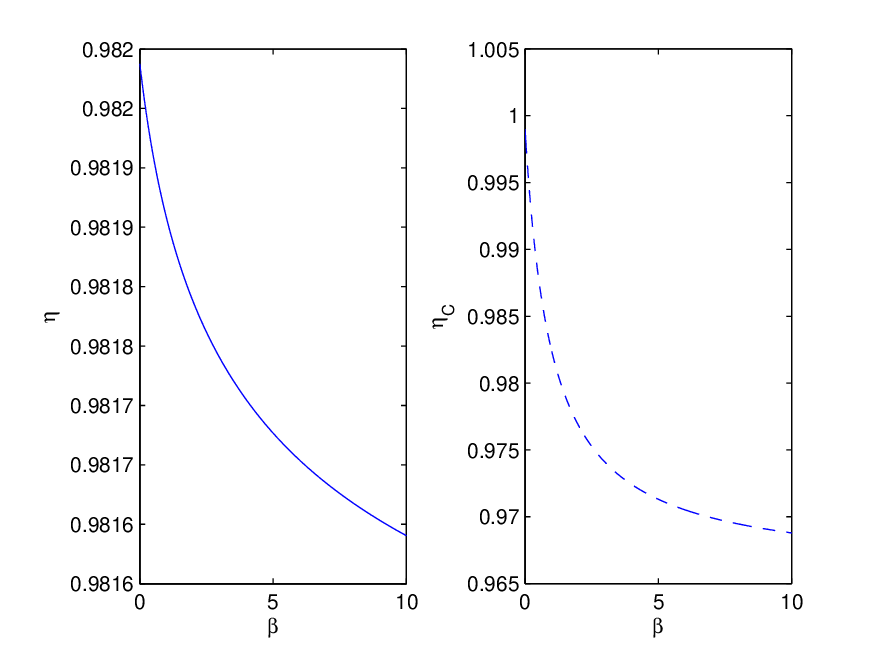}
\caption{\label{fig.2} The black hole heat efficiencies for a cycle with two isobars and two isochores/adiabats (left panel) for $q=1$, $r_1=1$, $r_2=10$, $l_1=1$, $l_2=10$ and two isotherms and two isochores/adiabats (right panel for Carnot's cycle) for $q=1$, $l_1=1$, $l_2=10$, $r_1=1$, and $r_2=10$.}
\end{figure}
The $T_H$ corresponds to $P_1$ and the $T_C$ corresponds to $P_2$ within EoS, Eq. (9).
Figure 2 shows that the black hole heat efficiency $\eta$ for a cycle with two isobars and two isochores/adiabats is less than $\eta_C$ for two isotherms and two isochores/adiabats for a Carnot cycle (for some range of coupling $\beta$).
 According to Fig. 2 when coupling $\beta$ increases the black hole heat efficiencies $\eta$ and $\eta_C$ decrease. The ratio $\eta/\eta_C$ is plotted in Fig. 3.
 \begin{figure}[h]
\includegraphics [height=2.7in,width=5in] {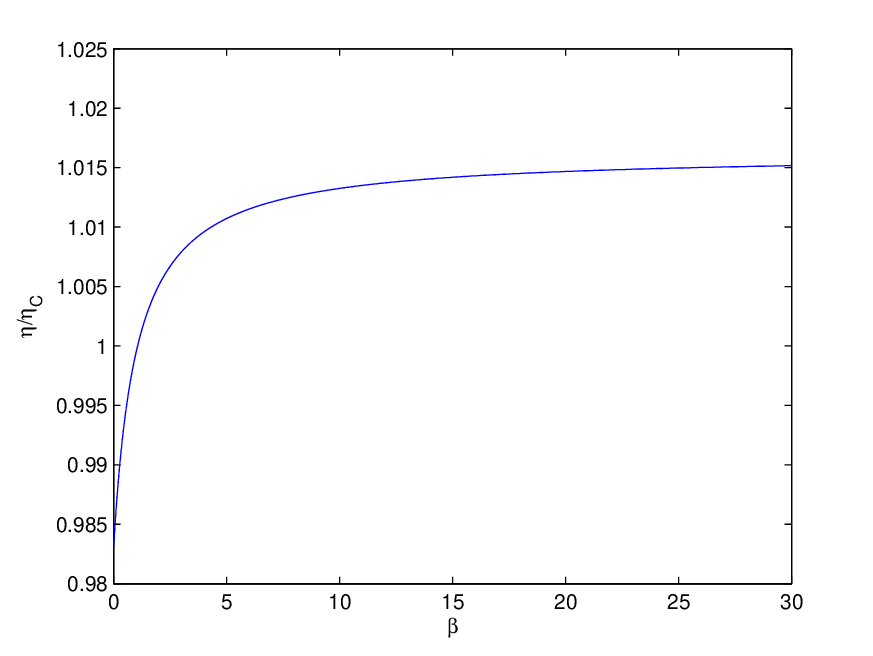}
\caption{\label{fig.3} The ratio $\eta/\eta_C$ vs. $\beta$  for $q=1$, $r_1=1$, $r_2=10$, $l_1=1$, $l_2=10$. For $\beta>1$  we have nonphysical case as $\eta>\eta_C$.}
\end{figure}
Figure 3 shows that at some range of coupling $\beta$ we have $\eta<\eta_C$, but for grater $\beta$ one has $\eta>\eta_C$
(the nonphysical case). This can be explained by the fact that RNED becomes Maxwell electrodynamics at the limit $\beta\rightarrow 0$. Therefore,
the physical case takes place only for small $\beta$. When $\beta\rightarrow 0$ we come to RN-AdS spacetime.
The ratio $\eta/\eta_C$ monotonously increases with the increase of coupling $\beta$ and for large $\beta$ the relation $\eta>\eta_C$  holds
that contradicts the traditional thermodynamics that the Carnot heat engine possesses the highest efficiency which is also nonphysical case. Making use of Eqs. (10) and (13) we depicted the black hole heat efficiencies $\eta$ and $\eta_C$ vs. $q$ in Fig. 4.
\begin{figure}[h]
\includegraphics [height=2.7in,width=5in] {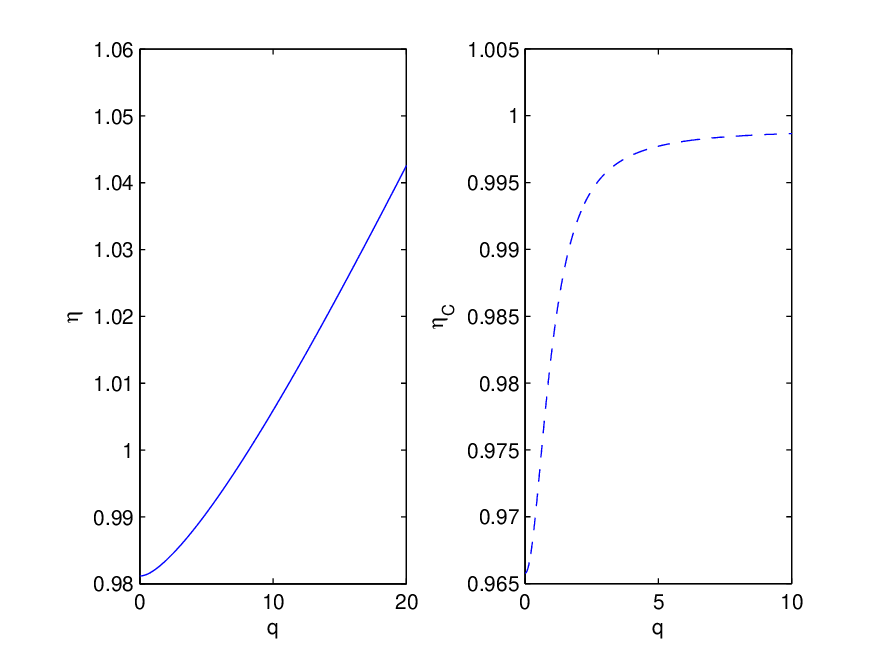}
\caption{\label{fig.4}  Left panel: The black hole heat efficiencies for a cycle with two isobars and two isochores/adiabats for $\beta=1$, $r_1=1$, $r_2=10$, $l_1=1$, $l_2=10$. Right panel for Carnot's cycle: Heat efficiency of two isotherms and two isochores/adiabats for $\beta=1$, $r_1=1$, and $r_2=10$, $l_1=1$, $l_2=10$.}
\end{figure}
In accordance with Fig. 4 if magnetic charge $q$ increases the black hole heat efficiencies $\eta$ and $\eta_C$ also increase. For Carnot's cycle (right panel) the black hole heat efficiency is bounded, $\eta_C<1$ showing the conversion of heat into work.  Left panel of Fig. 4 shows two regimes: 1) when $\eta \leq 1$ there is  the conversion of heat into work (the physical case); 2) if $\eta > 1$ we have creating a temperature difference through the application of work to the system which is probably nonphysical case. One can speculate that this is a refrigeration system - the work is produced by surrounds. But for large charge $q$ and small $\beta$ the temperature becomes negative (see Eq. 8) and, therefore, the system is nonphysical (see \cite{Johnson2} for the case of RN-AdS spacetime with the electrical charge).

These two regimes are separated by equation $\eta=1$ ($Q_C=0$)
\begin{equation}
\frac{r_2-r_1}{2}+\frac{r_2^3-r_1^3}{2l_2^2}-\frac{q^{3/2}\left(g(r_2)-g(r_1)\right)}{8\sqrt{2}\beta^{1/4}}=0.
\label{14}
\end{equation}
This equation depends on many parameters: $V_1$, $V_2$ (or $r_1$, $r_2$), $q$, $\beta$ and $l_2$ (or $P_2$) (see the rectangle cycle in Fig. 1).
In figures 1-3 we have considered just an example for a particular choice of parameters of a cycle. The solution to Eq. (14) for parameters given in Fig. 3 is $q\approx 8.1857$. It follows from Eq. (10) that the efficiency of the heat engine $\eta>1$ when $Q_H$ or $Q_C$ is negative  which corresponds to a refrigeration system. One can verify that for parameters $\beta=1$, $r_1=1$, and $r_2=10$, $l_1=1$, $l_2=10$ in Fig. 4, we have $Q_H\approx 474.34$ and $Q_C\approx -20.17$ for $q=20$ and $\eta=1.043$ that is the refrigeration system.
It is worth noting that Johnson \cite{Johnson2} demonstrated that the black hole heat engine at the critical point approaches the Carnot cycle when the electric charge increases. It was described there how to choose the physical cycle which touches the critical point. The same procedure can be used for our model by taking into account critical points found in \cite{Kruglov2}.

\section{Cycle in the vicinity of a critical point}

Firstly, the cycles of holographic heat engines close to critical points were considered in \cite{Johnson2}. Johnson showed that an efficiency of holographic heat engine of a cycle in the vicinity of this critical point approaches the Carnot efficiency. Here, we will study holographic heat engines cycle in the vicinity of a critical point for our model. We demonstrate the limit $\eta\rightarrow \eta_C$ as expected due to general universality for a black hole holographic heat engine near critical points \cite{Mann2}. The sub-leading term in an efficiency of a heat engine
includes the van der Waals ratio $Tc/(v_cP_c)=8/3$ ($v_c$ is the critical specific volume in our model) plus the term with RNED coupling and the parameters $x<1$, $y<1$ ($\Delta V=xV_c$, $\Delta P=yP_c$) \cite{Mann2}. It was shown in \cite{Mann2} that near critical point expansion of any models is given by
\begin{equation}
\rho=A\tau+B_0\tau\omega^{\delta-1/\beta}+C_0\omega^\delta+...,
\label{15}
\end{equation}
where parameters $\rho$, $\tau$ and $\omega$ are connected with the pressure $P$, temperature $T$ and volume $V$ as $P=P_c(\rho+1)$, $T=T_c(\tau+1)$, $V=V_c(\omega+1)$. \footnote{In our notations \cite{Kruglov2} $\rho\equiv p-1$, $p=P/P_c$, $\tau\equiv t$, $B=-B_0$, $C\equiv C_0$. } Taking into account the values of critical exponents in our model \cite{Kruglov2} $\delta=3$, $\beta=1/2$ (do not mix this $\beta$ with the RNED coupling $\beta$), we obtain (in our notations)
\begin{equation}
p=\rho+1=1+At-Bt\omega+C\omega^3+...,
\label{16}
\end{equation}
where (see \cite{Kruglov2})
\begin{equation}
A=\frac{1}{\rho_c}=\frac{8}{3}-\frac{2}{27q^2}\beta+{\cal O}(\beta^2),~~\rho_c=\frac{3}{8}+\frac{\beta}{96q^2}+
{\cal O}(\beta^2),~~B=\frac{1}{3\rho_c}=\frac{1}{3}A.
\label{17}
\end{equation}
Then, making use of the general expression for the efficiency of the cycle $\eta$ in the $x\rightarrow 0$ limit \cite{Mann2}, we obtain
\[
\eta=\frac{y}{A+y}-\frac{yB_0}{(A+y)^2(\delta-1/\beta+1)}(-x)^{\delta-1/\beta}+{\cal O}(x^{2\delta-2/\beta})
\]
\begin{equation}
=\frac{y}{A+y}-\frac{xyA}{6(A+y)^2}+{\cal O}(x^2).
\label{18}
\end{equation}
Note that the heat engine efficiency (18) depends only on the parameter $A$ which includes RNED coupling $\beta$. The Carnot efficiency $\eta_C =1-T_C/T_H$ is given by \cite{Mann2}
\begin{equation}
\eta_C =\frac{y}{A+y}+\frac{C_0}{A+y}(-x)^\delta+...=\frac{y}{A+y}+{\cal O}(x^3),
\label{19}
\end{equation}
where we have used the value $\delta=3$. Equations (18) and (19) show that $\eta/\eta_C<1$ as it was expected. Making use of Eqs. (18) and (19 we obtain the ratio
\begin{equation}
\frac{\eta}{\eta_C} =1-\frac{xA}{6(A+y)}+{\cal O}(x^3)\approx 1-\frac{x}{6}+\frac{xy}{16}+\frac{xy\beta}{576q^2}+{\cal O}(x^3),
\label{20}
\end{equation}
where the forth term in Eq. (20) is very small in our approximation ($xy\beta\ll xy$). Equation (20) shows that heat efficiency $\eta$ approaches to the Carnot efficiency and is defined mostly by parameter $x$ because $xy\ll x$. The expression (20) is close to Johnson’s formula for the ratio of efficiency $\eta$ to the Carnot efficiency $\eta_C$ in the vicinity of critical points of charged AdS black holes.

\section{Summary}

We have studied the heat engine of magnetically charged black holes as the working substance within Einstein's gravity in AdS space coupled to rational nonlinear electrodynamics. The dynamical negative  cosmological constant in extended phase space is treated as a thermodynamic pressure and black hole mass as an enthalpy. The efficiencies of black hole heat engines for rectangle closed path in the $P-V$ plane and for Carnot cycle were obtained. We have demonstrated that the black hole efficiencies for both cycles decrease when the nonlinear electrodynamics coupling $\beta$ increases and the black hole efficiencies increase if the magnetic charge increases. For small coupling $\beta$ we have the physical case $\eta<\eta_C$ but for grater $\beta$ one has nonphysical case $\eta>\eta_C$. For the Carnot cycle the black hole efficiency $\eta_C$ is bounded, $\eta_C<1$, but for rectangle closed path the efficiency grows with magnetic charge $q$ showing that when $\eta<1$ we have work done by black holes and if $\eta\geq 1$ one has a refrigeration system - the work is produced by surrounds (the nonphysical case as the temperature can be negative). An efficiency of the holographic heat engine of a cycle in the vicinity of a critical point was studied. Our result is similar to Johnson's result \cite{Johnson2} for electrically charged AdS black holes. One can employ the benchmarking scheme of \cite{Johnson4} to compare the heat efficiency of our model with heat efficiencies of other models. It should be noted that the stability analysis
of black holes is important for models to be viable. We leave such study for further. It is of interest also to study how quantum corrections can influence the thermodynamic properties and efficiency of the black hole heat engine. We will study this in further.
Our model of black hole physics in AdS space can be of interest for the study of gauge theories through holography.\\

\textbf{Data Availability Statement}: No Data associated in the manuscript.\\

\textbf{Competing interests}: The authors declare there are no competing interests.


\begin{thebibliography}{99}

\bibitem{Kastor} D. Kastor, S. Ray, and J. Traschen, Enthalpy and the Mechanics of AdS Black Holes, Class. Quant. Grav. \textbf{26}, 195011 (2009).
\bibitem{Dolan1} B. Dolan, The cosmological constant and the black hole equation of state, Class. Quant. Grav. \textbf{28}, 125020 (2011).
\bibitem{Dolan2} B. P. Dolan, Pressure and volume in the first law of black hole thermodynamics, Class. Quant. Grav. \textbf{28}, 235017 (2011).
\bibitem{Dolan3} B. P. Dolan, Compressibility of rotating black holes, Phys. Rev. D \textbf{84}, 127503  (2011).
\bibitem{Cvetic2} M. Cvetic, G. Gibbons, D. Kubiznak, and C. Pope, Black Hole Enthalpy and an Entropy Inequality for the
Thermodynamic Volume, Phys. Rev. D \textbf{84}, 024037 (2011).
\bibitem{Mann} S. Gunasekaran, R. B. Mann and D. Kubiznak, Extended phase space thermodynamics for charged and rotating black holes and Born—Infeld vacuum polarization, JHEP \textbf{1211}, 110  (2012).    
\bibitem{Maldacena} J. M. Maldacena, The Large N limit of superconformal field theories and supergravity, Int. J. Theor. Phys. \textbf{38}, 1113-1133 (1999); Adv. Theor. Math. Phys. \textbf{2}, 231 (1998).
\bibitem{Witten} E. Witten, Anti-de Sitter space and holography, Adv. Theor. Math. Phys. \textbf{2}, 253-291  (1998).
\bibitem{Witten1} E. Witten, Anti-de Sitter space, thermal phase transition, and confinement
in gauge theories, Adv. Theor. Math. Phys. \textbf{2}, 505-532 (1998).
\bibitem{Kovtun}P. Kovtun, D. T. Son and A. O. Starinets, Viscosity in strongly interacting quantum field theories from black hole physics, Phys. Rev. Lett. \textbf{94}, 111601 (2005).
\bibitem{Kovtun1}S. A. Hartnoll, P. K. Kovtun, M. Muller and S. Sachdev, Theory of the Nernst effect near quantum phase transitions in condensed matter, and in dyonic black holes, Phys. Rev. B \textbf{76}, 144502 (2007).
\bibitem{Hartnoll} S. A. Hartnoll, C. P. Herzog and G. T. Horowitz, Building a Holographic Superconductor, Phys. Rev. Lett. \textbf{101}, 031601  (2008).
\bibitem{Johnson} C. V. Johnson, Holographic Heat Engines, Class. Quant. Grav. \textbf{31}, 205002 (2014).
\bibitem{Johnson3} C. V. Johnson, Born--Infeld AdS black holes as heat engines, Class. Quant. Grav. \textbf{33}, 135001 (2016).
\bibitem{Mann1} R. A. Hennigar, F. McCarthy, A. Ballon, and  R. B. Mann, Holographic heat engines: general considerations and rotating black holes, Class. Quant. Grav. \textbf{34}, 175005 (2017).    
\bibitem{Johnson4}Avik Chakraborty and Clifford V. Johnson, Benchmarking black hole heat engines, I, Int. J. Mod. Phys. D \textbf{27} 16, 1950012 (2018).    
\bibitem{Johnson7}A. Chakraborty and C. V. Johnson, Benchmarking Black Hole Heat Engines, II, Int. J. Mod. Phys. D \textbf{27}, 1950006 (2018).
\bibitem{Johnson5} Clifford V. Johnson and Felipe Rosso,  Holographic Heat Engines, Entanglement Entropy, and Renormalization Group Flow,
Class. Quant. Grav. \textbf{36}, 015019 (2019).
\bibitem{Pedraza}Elena Caceres, Phuc H. Nguyen, Juan F. Pedraza, Holographic entanglement entropy and the extended phase structure of STU black holes, JHEP \textbf{1509},  184 (2015).
\bibitem{Johnson6} Clifford V. Johnson, Taub-Bolt Heat Engines, Class. Quant. Grav. \textbf{35}, 045001 (2018).
\bibitem{Zhang} J. Zhang, Y. Li, and H. Yu, Eur. Phys. J. C \textbf{78}, 645 (2018).
\bibitem{Rosso} F. Rosso, Holographic heat engines and static black holes: a general efficiency formula, Int. J. Mod. Phys. D \textbf{28}, 02 (2018).
\bibitem{Yerra}P. K. Yerra and B. Chandrasekhar, Heat engines at criticality for nonlinearly charged black holes, Mod. Phys. Lett. A \textbf{34}, 1950216 (2019).
\bibitem{Bamba}H. Ghaffarnejad, E. Yaraie, M. Farsam, and K. Bamba, A Note on Critical Nonlinearly Charged Black Holes,
 Nucl. Phys. B \textbf{952}, 114941 (2020).
\bibitem{Johnson8} C. V. Johnson, Holographic Heat Engines as Quantum Heat Engines, Class. Quant. Grav. 37, 034001 (2020).
\bibitem{Ahmed}W. Ahmed, H. Z. Chen, E. Gesteau, R. Gregory, and A. Scoins, Conical Holographic Heat Engines, Class. Quant. Grav. \textbf{36},
214001 (2019).
\bibitem{Yerra1} P. K. Yerra and C. Bhamidipati, Critical heat engines in massive gravity, Class. Quant. Grav. \textbf{37}, 205020 (2020).
\bibitem{Moumni} H. El Moumni and K. Masmar, Regular AdS black holes holographic heat engines in a benchmarking scheme, Nucl. Phys. B \textbf{973}, 115590 (2021).
\bibitem{Jafarzade} Kh. Jafarzade and J. Sadeghi, Phase transition and holographic in modified Horava–Lifshitz black hole, Int. J. Mod. Phys. D \textbf{\textbf{26}}, 1750138 (2017).
\bibitem{Ujjal} Ujjal Debnath, Higher-dimensional polytropic and modified Chaplygin black holes: Thermodynamics and heat engines,
 Mod. Phys. Lett.A \textbf{37}, 2250243 (2022).
\bibitem{Kruglov}S. I. Kruglov, A model of nonlinear electrodynamics, Ann. Phys. \textbf{353}, 299-306 (2014).
\bibitem{Kruglov1} S. I. Kruglov, Remarks on Nonsingular Models of Hayward and Magnetized Black Hole with Rational Nonlinear Electrodynamics,  Grav. Cosmol. \textbf{27}, 78-84  (2021).    
\bibitem{Bronnikov} K. A. Bronnikov, Regular magnetic black holes and monopoles from nonlinear electrodynamics, Phys. Rev. D \textbf{63}, 044005 (2001).
\bibitem{Kruglov2} S. I. Kruglov, Rational non-linear electrodynamics of AdS black holes and extended phase space thermodynamics, Eur. Phys. J. C \textbf{82}, 292 (2022).
\bibitem{Behnam}B. Pourhassan, S. H. Hendi, S. Upadhyay, \.{I}. Sakallı, and E. N. Saridakis, Thermal fluctuations of (non)linearly charged BTZ black hole in massive gravity,  Int. J. Mod. Phys, D \textbf{32}, 2350110 (2023).
\bibitem{Behnam1} B. Pourhassan and \.{I}. Sakallı, Non-perturbative correction to the Ho\u{r}ava–-Lifshitz black hole thermodynamics,
 Chin. J. Phys. \textbf{79}, 322-338 (2022).
\bibitem{Bronnikov2} K. A. Bronnikov, In: Regular magnetic black holes. Towards a New Paradigm of Gravitational Collapse.
 Cosimo Bambi Editor. Springer Series in Astrophysics and Cosmology. pp. 37-66 (2001), https://doi.org/10.1007/978-981-99-1596-5.
\bibitem{Kruglov5}S. I. Kruglov, On generalized Born-Infeld electrodynamics, J. Phys. A \textbf{43}, 375402 (2010).
\bibitem{Kruglov6}S. I. Kruglov, Notes on Born–Infeld-type electrodynamics, Mod. Phys. Lett. A \textbf{32}, 1750201 (2017).
\bibitem{Kruglov7}S. I. Kruglov,  Nonlinear electrodynamics with birefringence, Phys. Lett. A \textbf{379}, 623-625 (2015).
\bibitem{Kruglov3}  S. I. Kruglov, Universe acceleration and nonlinear electrodynamics, Phys. Rev. D \textbf{92}, 123523 (2015).
\bibitem{Kruglov4} S. I. Kruglov, The shadow of M87* black hole within rational nonlinear electrodynamics, Mod. Phys. Lett. A \textbf{35}, 2050291 (2020).   
\bibitem{Hawking}  S. W. Hawking and D. N. Page, Thermodynamics of black holes in anti-de Sitter space, Commun. Math. Phys. \textbf{87}, 577 (1983).
\bibitem{Johnson1} C. V. Johnson, An Exact Efficiency Formula for Holographic Heat Engines, Entropy \textbf{18}, 120 (2016).
\bibitem{Johnson2} C. V. Johnson, Exact model of the power-to-efficiency trade-off while approaching the Carnot limit, Phys. Rev. D \textbf{98}, 026008  (2018).
\bibitem{Mann2}Maria C. DiMarco, Sierra L. Jess, Robie A. Hennigar, and Robert B. Mann, Universality for black hole heat engines near critical points, Phys. Rev. D \textbf{107}, 044001 (2023).

\end{thebibliography}
\end{document}